\documentstyle[12pt,WSPRL]{article}

\begin{document}

\centerline{\normalsize\bf BILEPTON RESONANCE IN ELECTRON-ELECTRON SCATTERING} 

\baselineskip=16pt
\vspace*{0.3cm}

\centerline{\footnotesize PAUL H. FRAMPTON}
\baselineskip=13pt
\centerline{\footnotesize\it Department of Physics and Astronomy,}
\baselineskip=12pt
\centerline{\footnotesize\it University of North Carolina, Chapel Hill, NC 27599-3255}
\centerline{\footnotesize E-mail: frampton@physics.unc.edu}
\vspace*{0.9cm}

\abstracts{ 
Theoretical backgound for bileptonic gauge bosons is reviewed, both
the SU(15) GUT model and the 3-3-1 model.  
Mass limits on bileptons are discussed coming from $e^+e^-$ scattering, polarized
muon decay and muonium-antimuonium conversion. 
Discovery in $e^-e^-$ at a linear collider at low energy (100GeV)
and high luminosity ($10^{33}/cm^2/s$) is emphasised.}

\bigskip
\bigskip
\noindent

\bigskip
{\bf Introduction.}

\bigskip
\bigskip

It is a stunning historical fact that $e^-e^-$ collisions
have never been studied at a center of mass
energy above 1.12 GeV as published in 1971 by
Richter {\it et al}\cite{BOGR}. There were plans to explore $e^-e^-$
at DESY but these were abandoned when money ran out.

\bigskip
\bigskip

The three large projects in HEP for the US (and internationally)
for the foreseeable future are: NLC, MC and VLHC. 
Of these the NLC is for the first decade of the twenty-first
century; the other two are for the second decade. The NLC is presently a multi-
billion dollar project primarily aimed at $e^+e^-$. 

\bigskip
\bigskip

A topic of
this workshop is: should it have also $e^-e^-$ capability?

\bigskip
\bigskip

Why has $e^-e^-$ been so neglected? Firstly $e^+e^-$ is
where $Z^{'}$ can be found - often cited as the most conservative
extension of the Standard Model (SM). By contrast $e^-e^-$ is
an exotic, empty channel because it has double electric
charge and lepton number $L=2$. Surely, $e^-e^-$ would
allow only checks of higher-order quantum electrodynamics.
But physics is an experimental science!

\bigskip
\bigskip

{\bf $e^-e^-$ Resonance.}

\bigskip
\bigskip

Such a resonance must have $L=2$ and $Q=2$. It must be a boson.
For spin zero a doubly-charged Higgs scalar, the coupling is a free parameter
and is generically small. For a spin one gauge boson, the coupling
is large and prescribed. Bilepton gauge bosons give a pronounced peak
at $s = M^2$. But, as our main emphasis here, the resonance tail
is detectable at much lower energy.

\bigskip
\bigskip

Bilepton gauge bosons were first suggested in the context of SU(15) grand 
unification\cite{FL}.

\bigskip
\bigskip

First recall that in $SU(5)$ grand unification with families each in $5 + 
\bar{10}$
the reason for $B$ violation is that the second rank tensor $\bar{10}$ has
indefinite $B$ and $L$ quantum numbers.

\bigskip

If $SU(5)$ had fermions only in the $5$ then $B$ and $L$ would necessarily be 
conserved perturbatively.

\bigskip
\bigskip

The presence of the $\bar{10}$ is what causes the indeterminacy of
$B$ and $L$ and allows mediation of proton decay in the gauge sector.

\bigskip
\bigskip

Since proton decay remains elusive the idea in $SU(15)$ is to prohibit it in
the gauge sector. The 15 helicity states in each family are assigned
to a $15$ of $SU(15)$. Whereupon each gauge boson has definite $B$ and $L$ 
according
to which pair of the fundamental fermions it couples.

\bigskip
\bigskip

The first family is assigned to:

\bigskip
\bigskip

\[ 15 = (u_L^R, u_L^G, u_L^B, d_L^R, d_L^G, d_L^B; \bar{u}_L^R, 
\bar{u}_L^G, \bar{u}_L^B,\bar{d}_L^R, \bar{d}_L^G, \bar{d}_L^B; e_L^+, \nu_{eL}, e_L^- ) \]

\bigskip
\bigskip

\noindent and similarly for the second and third families.

\bigskip
\bigskip

\noindent It is clear that all of the 224 gauge bosons of $SU(15)$ have definite
$B$ and $L$.

\bigskip 
\bigskip

Anomaly cancellation is by mirror fermions - disfavored aesthetically but not 
phenomenologically.

\bigskip
\bigskip

The pattern of spontaneous symmetry breaking is:

\bigskip
\bigskip

\[ SU(15) \stackrel{M_G}{\longrightarrow} SU(12)_q \times SU(3)_l \]

\[ ~~~~~~~~~~~~~~~~~~~~~~~~  \stackrel{M_B}{\longrightarrow} SU(6)_L \times 
SU(6)_R \times U(1)_B \times SU(3)_l \]

\[ ~~~~~~~~~~~~~~~ \stackrel{M_A}{\longrightarrow} SU(3)_C \times SU(2)_L \times
U(1)_Y \]


\bigskip
\bigskip

\noindent  In the breaking at $M_A$ color $SU(3)_C$ is embedded in $SU(6)_L 
\times SU(6)_R$
as $(3 + 3, 1) + (1, \bar{3} + \bar{3})$.

\bigskip
\bigskip

\noindent $SU(2)_L$ is embedded in $SU(6)_L \times SU(3)_l$ with $6_L = 3(2)_L$
and $3_L = 2_L + 1_L$

\bigskip
\bigskip

\noindent $U(1)_Y$ is contained in $SU(6)_R \times U(1)_B \times SU(3)_l$ according to:

\bigskip
\bigskip

\[ Y = \sqrt{3} \Lambda + \sqrt{\frac{2}{3}} B + \sqrt{3} {\cal Y}  \]

\bigskip
\bigskip
\bigskip

\noindent with $\Lambda$, $B$ and ${\cal Y}$ generators of $SU(6)_R$, $U(1)_B$ 
and $SU(3)_l$, respectively, normalized as $SU(15)$ matrices with

\bigskip
\bigskip

\[ Tr (\Lambda^a \Lambda^b ) = 2 \delta^{ab}. \]

\bigskip
\bigskip
\bigskip

\noindent  Explicitly, these normalized $SU(15)$ generators are

\bigskip
\bigskip

\[ \Lambda = \frac{1}{\sqrt{3}} {\rm diag} ( 000000, -1-1-1 111, 000) \]

\bigskip

\[  B = \sqrt{\frac{3}{2}} {\rm diag} (\frac{1}{3} \frac{1}{3} \frac{1}{3}
\frac{1}{3} \frac{1}{3} \frac{1}{3}, -\frac{1}{3} -\frac{1}{3}-\frac{1}{3} 
-\frac{1}{3} -\frac{1}{3} -\frac{1}{3}, 000 ) \]

\bigskip
\bigskip

\[ {\rm and}~~{\cal Y} =  \frac{1}{\sqrt{3}} {\rm diag} ( 000000, 000000, 2 -1 
-1) \]

\bigskip

\noindent RENORMALIZATION GROUP

\bigskip
\bigskip

\[ \mu d\alpha_i(\mu) / d\mu = B_i \alpha_i^2(\mu) \]

\bigskip
\bigskip

\noindent with matching conditions, at $M_A$:

\[ \alpha_{3C}^{-1}(M_A) = \frac{1}{2} \alpha_{6L}^{-1}(M_A) + \frac{1}{2} 
\alpha_{6R}^{-1} (M_A) \]

\bigskip

\[ \alpha_{2L}^{-1}(M_A) = \frac{3}{4} \alpha_{6L}^{-1}(M_A) + \frac{1}{4} 
\alpha_{3l}^{-1} (M_A) \]

\bigskip

\[ \alpha_{1Y}^{-1}(M_A) = \frac{9}{20} \alpha_{6R}^{-1}(M_A) + \frac{1}{10} 
\alpha_{B}^{-1} (M_A) + \frac{9}{20} \alpha_{3l}^{-1}(M_A) \]

\bigskip

\noindent at $M_B$:

\bigskip

\[ \alpha_{6L}(M_B) = \alpha_{6R}(M_B) = \alpha_B(M_B) = \alpha_{12q}(M_B) \]

\bigskip

\noindent and at $M_G$:

\[ \alpha_{12q}(M_G) = \alpha_{3l}(M_G) = \alpha_{15}(M_G) \]

\bigskip
\bigskip

\noindent The results can be tabulated, as shown in this  
Table of typical values for the three breaking scales of $SU(15)$

\bigskip
\bigskip

\begin{center}

\begin{tabular}{||c|c|c||}    \hline\hline
$M_A(GeV)$ & $M_B(GeV)$ & $M_G(GeV)$ \\ \hline
$250$ & $4.0 \times 10^6$ & $6.0 \times 10^6$ \\
$500$ & $5.8 \times 10^6$ & $8.9 \times 10^6$ \\
$10^3$ & $8.3 \times 10^6$ & $1.3 \times 10^7$ \\
$2 \times 10^3$ & $1.2 \times 10^7$ & $1.9 \times 10^7$ \\  \hline\hline
\end{tabular}

\end{center}

\bigskip
\bigskip

\noindent There is one input parameter, say $M_A$. 

\bigskip

\noindent $M_B$ and $M_G$ are outputs.

\bigskip
\bigskip

\noindent At low energies ($M_A$) the gauge bosons under $SU(6)_L \times SU(6)_R
\times U(1)_B \times SU(3)_l$ are, with respect to the standard model:

\bigskip
\bigskip
\
\[ 35_L = (8,3)_0 + (8,1)_0 + (1,3)_0 \]

\bigskip
\bigskip

\[ 35_R = 2(8,1)_0 + (8,1)_{\pm 1} + (1, 1)_0 + (1, 1)_{\pm 1} \]

\bigskip

\[ 1_B = (1, 1)_0 \]

\bigskip

\[ 8_l = (1, 3)_0 + (1, 1)_0 + (1, 2)_{\pm 3/2} \]

\bigskip

\noindent All are interesting but the last-listed $(1, 2)_{\pm 3/2}$ are the 
bileptonic gauge bosons which can show up in Moller scattering.

\noindent ( {\it e.g.} $e^-e^- \longrightarrow \mu^-\mu^-$).

\bigskip

\noindent Clearly such bileptons are a general feature  

\noindent of the embedding

\bigskip

\[ SU(2)_L \subset SU(3) \]

\bigskip

\noindent and have the electric charges 

\bigskip

\noindent $(Y^{--}, Y^{-})$ ~~ (L = +2) 

\bigskip

\noindent with antiparticles 

\bigskip

\noindent $(Y^{++}, Y^{+})$ ~~ (L = -2).

\bigskip
\bigskip

\noindent This feature of $SU(15)$ grand unification re-emerges in the $3-3-1$ 
model\cite{PHF} to which we now turn.

\bigskip
\bigskip

\noindent $3-3-1$ is more economic, and anomaly cancellation is more elegant, 
compared to $SU(15)$.

\bigskip

\noindent To introduce the 3-3-1 model, the following are motivating factors:

\bigskip

\noindent 1. Consistency of a gauge theory requires cancellation of all chiral 
anomalies. Such cancellation occurs for a quark-fermion family and is enough 
(almost) to fix all charges.

\bigskip

\noindent 2. This does not explain $N_f > 1$ but is sufficiently impressive to 
suggest that $N_f = 3$ may be explicable by anomaly cancellation in an 
extension. This requires extended families have non-zero anomaly and not all 
three families treated similarly.

\bigskip

\noindent 3. The third family is exceptional because of the top quark mass, and 
suggests 

 +1 +1 -2 cancellation.

\bigskip

\noindent 4. There is such a -2 in the SM as the ratio of quark charges.

\bigskip

\noindent 5. Extension of $SU(2)_L$ to $SU(3)_L$ will have the same {\it lepton}
couplings of the bileptons as in $SU(15)$.

\bigskip
\bigskip

\noindent For the 3-3-1 model the gauge group is:

\bigskip

\noindent $SU(3)_C \times SU(3)_L \times U(1)_X$
 
\bigskip

\noindent The first family quarks are assigned to

\bigskip

\[ \left( \begin{array}{c} u \\ d \\ D \end{array} \right)_L
~~~ \bar{u}_L~~~ \bar{d}_L~~~\bar{D}_L \]

\bigskip

\noindent The triplet is a 3 of $SU(3)_L$.

\bigskip

\noindent The second family of quarks is assigned similarly:

\bigskip

\[ \left( \begin{array}{c} c \\ s \\ S \end{array} \right)_L
~~~ \bar{c}_L~~~ \bar{s}_L~~~\bar{S}_L \]

\bigskip

\noindent The third family of quarks is assigned differently: 

\bigskip

\[  \left( \begin{array}{c} T \\ t \\ b \end{array} \right)_L
~~~ \bar{T}_L~~~ \bar{t}_L~~~\bar{b}_L \]

\bigskip
\bigskip

\noindent The triplet in this case is a 3* of $SU(3)_L$.

\bigskip

\noindent The X quantum numbers of the triplets are equal to the electric 
charges of the cental members. That is, for the three families of quarks, 
$X = -\frac{1}{3},~~-\frac{1}{3},~~+\frac{2}{3}$.

\bigskip
\bigskip

\noindent The leptons are assigned to 3*'s as follows:

\bigskip
\bigskip

\[  \left( \begin{array}{c} e^{+} \\ \nu_e \\ e^{-} \end{array} \right)_L ~~~
\left( \begin{array}{c} \mu^{+} \\ \nu_{\mu} \\ \mu^{-} \end{array} \right)_L 
~~~
\left( \begin{array}{c} \tau^{+} \\ \nu_{\tau} \\ \tau^{-} \end{array} \right)_L
\]

\bigskip
\bigskip

\noindent  These three antitriplets have $X = 0$.

\bigskip
\bigskip

\noindent Let us see how anomalies cancel.
Recall that anomaly cancellation is crucial in many situations
of model-building beyond the standard model {\it e.g.} chiral color\cite{FG}
and in string theory\cite{FK}.

\bigskip
\bigskip

\noindent The color anomaly $(3_L)^3$ cancels because QCD is vectorlike. 

\bigskip
\bigskip

\noindent  The anomaly $(3_L)^3$ is non-trivial. Taking $N_C$ colors and $N_l$ 
light neutrinos
the anomaly cancels only if $N_C = N_l = 3$.

\bigskip
\bigskip

\noindent The remaining anomalies 

\bigskip

\noindent $(3_C)^2X,~(3_L)^2X,~X^3,~X(T_{\mu\nu})^2$ 

\bigskip

\noindent also all cancel.

\bigskip
\bigskip

\noindent In particular, each family has a non-zero anomaly for $X^3$, ~ 
$(3_L)^2X$ and $(3_L)^3$; in each case the anomalies cancel proportionately to 
$+1~+1~-2$, as anticipated in the earlier discussion.

\bigskip
\bigskip

\noindent To break the symmetry requires several Higgs multiplets.

\bigskip
\bigskip

\noindent First an $X=+1$ triplet $\Phi$ with VEV $<\Phi> = (0, 0, U)$ breaks 
331 to 321 and gives masses to the D, S and T quarks as well as the gauge bosons
$Z'$ and $Y$. The scale $U$ sets the scale for the new physics.

\bigskip
\bigskip

Electroweak symmetry breaking requires two further triplets $\phi$ and $\phi'$
with $X=0$ and $X=-1$ respectively. Their VEVs give mass to d, s, t and to u, c,
b respectively. The first VEV also gives a family-antisymmetric 
contribution to the charged leptons. 
To obtain a general mass matrix for charged leptons 
necessitates adding a sextet with $X = 0$.

\bigskip
\bigskip

\noindent THE NEW PHYSICS SCALE U

\bigskip
\bigskip

\noindent There is a lower bound from precision electroweak data:

\bigskip
\bigskip

\noindent $Z-Z'$ mixing dictates $M(Z')>300GeV$.

\bigskip
\bigskip

\noindent FCNC limits give a similar bound. For FCNC it is crucial that the 
third family be the one treated asymmetrically. Otherwise the FCNC disagree with
experiment.

\bigskip
\bigskip

\noindent  UPPER BOUND ON U:

\bigskip
\bigskip

\noindent A bound on $U$ arises because the embedding of 321 in 331 requires\cite{PHF} 
${\rm sin}^2\theta < 1/4$ because for ${\rm sin}^2\theta = 1/4$ the coupling 
$g_X$ diverges. This fixes $U<3TeV$ using ${\rm sin}^2\theta(M_Z)=0.231$.  Hence
$M(Y)$ cannot be higher than 1.5 TeV.

\bigskip
\bigskip

\noindent LEP data:

\bigskip
\bigskip

\noindent The highest precision high-energy data is from LEP. It gives\cite{FN}
$M(Y)>120GeV$.

\bigskip
\bigskip

\noindent The best lower bounds come from low energy experiments:

\bigskip
\bigskip

\noindent (1) Polarized muon decay\cite{muon}:

\bigskip
\bigskip

\noindent $M(Y^{\pm})>230$ GeV.

\bigskip
\bigskip

\noindent (2) Muonium-Antimuonium conversion\cite{willmann}:

\bigskip
\bigskip

\noindent $M(Y^{\pm\pm})>850GeV$.

\bigskip
\bigskip

\bigskip

\noindent Just to recapitulate some of the points made at the beginning:

$e^-e^-$ collisions have never been studied above c.o.m. energy 1.12 GeV.
An NLC should have $e^-e^-$ capability.

\bigskip
\bigskip

{\bf Accomplishment of $e^-e^-$ Collisions
at NLC.}

\bigskip
\bigskip

In the post-SSC era it is desirable to avoid a third comma
in the cost C, {\it i.e.} 
$C < \$1B$.

\bigskip
\bigskip

How can this be achieved?

\bigskip
\bigskip

The cost of an NLC is roughly linear in the energy.

\bigskip
\bigskip

A 500GeV NLC was costed last year at \$7.9B, although I have been told
informally\cite{Richter} that
that cost might be lowered 
below \$5B. Thus 100GeV could be below \$1B?

\bigskip
\bigskip
\bigskip

Therefore the first {\it fundable} step
could focus on luminosity rather than energy and be a 100GeV
machine with luminosity $\sim 10^{33}$. This is
sufficiently above LEP to give a Giga-Z 
and allows an opportunity to do
new machine physics.

\bigskip
\bigskip
\bigskip

{\bf Acknowledgements}

\bigskip
\bigskip

It is a pleasure to thank Clemens Heusch for organizing a pleasant meeting
and the US Department of Energy for support
under Grant No. DE-FG01-97ER41036.

\bigskip

\bigskip
\bigskip
\noindent {\bf Note Added}

\bigskip
\bigskip

In a recent work [P.H. Frampton and A. Rasin, UNC Report IFP-781-UNC (February 2000)]
we have updated the cross-section estimates for $e^-e^- \rightarrow \mu^-\mu^-$
in \cite{FN} which used the $SU(15)$ model. In the simpler 331-model the cross-section
is about one order of magnitude {\it higher} than the results in \cite{FN}.

\end{document}